\documentclass[conference]{IEEEtran}
\IEEEoverridecommandlockouts
\usepackage{cite}
\usepackage{amsmath,amssymb,amsfonts}
\usepackage{algorithm}
\usepackage{algorithmic}
\usepackage{graphicx}
\usepackage{textcomp}
\usepackage{amsmath}
\usepackage{svg}
\usepackage{xcolor}
\usepackage{comment}
\begin{document}
% \title{Mobile App Pentest Using Tramonto Framework\\
\title{Pentest on an Internet Mobile App: A Case Study using Tramonto*{\thanks{* This study was financed in part by the Coordenação de Aperfeiçoamento de Pessoal de Nível Superior - Brasil (CAPES) - Finance Code 001 and the Brazilian National Institute of Science and Technology in Forensic Sciences (INCT Forensic Sciences) project (Grant 465450/2014-8). Avelino F. Zorzo is supported by CNPq (grant 315192/2018-6). This paper was achieved in cooperation with HP Brasil using incentives of Brazilian Informatics Law (Law n 8.248 of 1991).  }}
}

\author{\IEEEauthorblockN{
Daniel Dalalana Bertoglio}
\IEEEauthorblockA{\textit{PUCRS and FEEVALE, Brazil} \\
dalalana@gmail.com}
\and
\IEEEauthorblockN{
Guilherme Girotto}
\IEEEauthorblockA{\textit{PUCRS, Brazil} \\
guilherme.girotto@edu.pucrs.br}
\and
\IEEEauthorblockN{
Charles Varlei Neu}
\IEEEauthorblockA{\textit{PUCRS and UNISC, Brazil} \\
charles.neu@edu.pucrs.br}
\and
\IEEEauthorblockN{
Roben Castagna Lunardi}
\IEEEauthorblockA{\textit{PUCRS and IFRS, Brazil} \\
roben.lunardi@edu.pucrs.br}
\and
\IEEEauthorblockN{
Avelino Francisco Zorzo}
\IEEEauthorblockA{\textit{PUCRS, Brazil} \\
avelino.zorzo@pucrs.br}
}

\maketitle

\begin{abstract}
Mobile applications are used to handle different types of data. Commonly, there is a set of personal identifiable information present in the data stored, shared and used by these applications. From that, attackers can try to exploit the mobile application in order to obtain or to cause private data leakage. Therefore, performing security assessments is an important practice to find vulnerabilities in the applications and systems before the application is deployed, or even during their use. Regarding security assessments, Penetration Test (Pentest) is one of the security test types that can be used to detect vulnerabilities through simulated attacks. Additionally, Pentest can be performed using different methodologies and best practices, through several frameworks to: organize the test execution, execute tools, provide estimations, provide reports and document a Pentest. One such framework is Tramonto, which aims to assist a cybersecurity expert during the Pentest execution by providing organization, standardization and flexibility to the whole Pentest process. This paper presents a Pentest case study applied to a Brazilian university Mobile App using the Tramonto framework. The main goal of this case study is to present how Tramonto can be applied during a Pentest execution, assisting cybersecurity experts in the tasks included in the Pentest process. Our results show details on how to perform a Pentest using Tramonto and the found vulnerabilities in the Mobile App. Besides that, there is a discussion about the main contributions obtained from our results, and we were able to verify that Tramonto managed, organized and optimized the whole Pentest process.
\end{abstract}


\section{Introduction}
 
Over the years, software-based systems have become part of our daily life. At the same time, these systems have got larger and more complex, increasing the possibilities to find vulnerabilities in those systems \cite{ZHANG200043}. In this sense, there is a need to better understand security practices and apply them during system development  \cite{hackinggrowth}. Several of those practices are shared by the security research community through experiences and case studies on security protection methods, processes or algorithms, applied to networks and systems  \cite{Gal-Or2004}. 

Several techniques and strategies to detect system vulnerabilities and to improve software security are available \cite{fbireport}. Penetration Test (Pentest) \cite{DALALANA2017}, is a security strategies that can be used to find vulnerabilities through simulated attacks through different methodologies and best practices. Furthermore, several frameworks can be used to organize the whole Pentest execution, manage tools, provide estimations, reports and documentation. One example of framework that can be used to perform Pentest is the Tramonto \cite{Tramonto2019} framework. Tramonto aims to assist a cybersecurity expert during the Pentest execution by providing organization, standardization and flexibility.

This paper presents an actual case study of a Pentest using Tramonto that aims to find vulnerabilities at a university Mobile Internet App. Tramonto helps to document and to describe the whole Pentest process\footnote{The guide explaining and showing how to use the framework is available on Tramonto's website - https://www.tramontosecurity.com}. All the sensitive data obtained during the Pentest (\textit{e.g.}, the university name) is omitted to avoid privacy issues. For this study, we used \textit{target.com.br} as the alias for the university website and \textit{Target University} as the university name. All the test process and the findings were reported to the security officers of the university. After that, the security issues were mitigated and properly treated - a new release of the Mobile App was produced in order to correct the breaches and weaknesses that were found during our Pentest.

The remaining of this paper is structured as follows. Section \ref{sec:background} presents the main concepts, definitions and subjects that are the base for this work - including a brief description of the Tramonto framework. Section \ref{sec:related} discusses related work. Section \ref{sec:specification} presents details about the case study and its specification. Section \ref{sec:execution} describes details about the attack vectors present in the Execution Step of the Tramonto framework - and also discusses the Pentest results. Section \ref{sec:lessons} presents the main contributions and lessons learned related to our research. Finally, Section \ref{sec:conclusion} concludes  this  paper  and  indicates  some
future work.

\section{Background - Pentest and Tramonto}
\label{sec:background}

Penetration test (Pentest) is a controlled tentative to pe\-netrate into a system or network in order to identify vulnerabilities \cite{DALALANA2017}. Pentest applies the same techniques that are used in a regular attack by a hacker to allow measures in order to eliminate vulnerabilities before they can be explored by unauthorized people. Usually, the Pentest process may be divided into the following activities: data gathering of the target system; scanning the target system to identify the available services/protocols; identifying existing systems and applications that are running on the target system; and identifying and exploit the known vulnerabilities on the systems and applications \cite{DALALANA2017}.

In order to improve systems security, several organisations, such as \textit{OWASP} \cite{MEUCCI2014}, introduced projects that aim to help people and companies to understand the most common vulnerabilities and mistakes, presenting examples on how to reproduce such attacks. One example that has helped companies to improve security in the past years is the \textit{OWASP Top Ten Project}~\cite{owasptop10}.

Despite of that, there are still challenges and open issues regarding to the manner that a Pentest is conducted, modelled and what methodology to use. Due to the variety of characteristics and functionalities, the main challenges faced by cybersecurity experts during the planning, or during the exploitation, are the lack of a standard way to use formal methods, models, methodologies and frameworks~\cite{DALALANA2017}.

The Tramonto framework \cite{Tramonto2019} was proposed as an alternative to help testers that needed flexibility. Tramonto aims to guide testers during a Pentest process according to three fundamental principles: \textbf{Organization}, to increase the planning capacity of the Pentest and management optimization, also allowing the tester to better understand the Pentest requirements; \textbf{Standardization}, to establish a common Pentest conduction structure and, at the same time, greater reliability considering the use of methods that are already known and applied by security professionals; and, \textbf{Flexibility}, which addresses the suitability of the tasks and concepts provided by Tramonto according to the tester experience. Tramonto also provides more flexibility to the whole Pentest conduction, although it was designed based on the following consolidated security test methodologies: OSSTMM \cite{HERTZOG2010}, ISSAF\cite{ISSAF2006}, PTES\cite{PTES2012}, OWASP\cite{MEUCCI2014} and NIST SP 800-15\cite{STOUFFER2008}.

\begin{figure}[h]
\centering
\includegraphics[width=0.7\columnwidth]{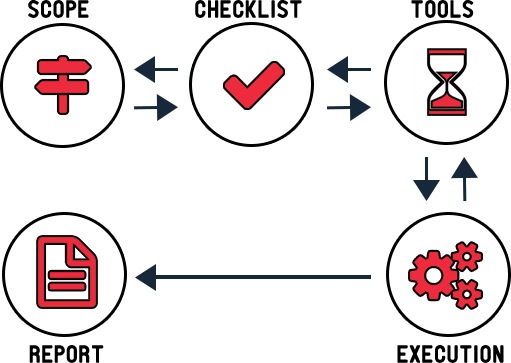}
\caption{Tramonto recommended steps.}
\label{fig:tramontosteps}
\end{figure}

Basically, Tramonto drives a Pentest through five steps: \textbf{1) Fitting Scope}, where  data management and initial choices about the scope and rules of engagement are initialized; \textbf{2) Performing Checklist}, to provide a checklist containing requirements, documents, artifacts and tasks for the Pentest plan; \textbf{3) Refinement Tools and Strategies}, as a place to inform the tester about possible strategies and tools that can be used in the Pentest; \textbf{4) Penetration Execution}, to describe the attack vectors and the actions performed to try the exploitation and other testing types; and \textbf{5) Final Findings and Report}, to prepare reports based on the previous information and artifacts obtained during the other steps. These steps are organized in order to provide a kind of script to the tester, allowing better management and control of activities. Nonetheless, any of the steps can be revisited at any time (see Figure \ref{fig:tramontosteps}).

\section{Related Work}
\label{sec:related}

This section describes some case studies on penetration test on specific institutions. These case studies use strategies that are very close related to the ones used in our work. 

Tiago Vieira and Carlos Serrão \cite{financial-portal}, for example, described a penetration test performed on a Financial Web Application. The authors had access to the institution internal network and used different techniques to discover a set of vulnerabilities, and analyzed a set of reports generated from Pentest executions over different Financial Web Applications. They used the \textit{OWASP Testing Guide (version 4)} methodology and the test was conducted under a controlled environment. The first task was to collect all existing vulnerabilities and categorize them by their risk as high, medium or low. The authors also used the \textit{OWASP Top Ten Project} to perform this risk segmentation and, finally, performed a risk analysis, which consisted of measuring the likelihood of a risk circumstance to occur and the potential impact of it.

Omeiza and Owusu performed a case study of a penetration test within an Educational Institute Portal \cite{educational-portal}. In order to perform the Pentest, the authors adopted an homemade test methodology to organize the Pentest process. The authors also classified the vulnerabilities into three levels (low, medium and high), according to their own methodology. Furthermore, the study discussed ways to mitigate and to fix all the security issues related to vulnerabilities that were found.

Both studies provide different approaches to perform security tests - specifically Pentest - regarding the importance to understand the characteristics, aspects, functionalities and how the tests were performed in different scenarios. Moreover, different methodologies to perform the tests were adopted, \textit{i.e.}, OWASP Testing Guide and a homemade  methodology. Our proposal uses the \textit{Tramonto} framework along with \textit{Tramonto-App} \cite{DALALANA2018} to assist the test conduction. Important to remember that through Tramonto any methodology can be used during the Pentest, hence, either works could have been performed using Tramonto. Other strategies to improve applications security could also be applied, for example, model-based testing\cite{peralta2008}. 

\section{Case Study - Pentest Specification}
\label{sec:specification}

The target scenario used in the case study, as mentioned previously, is represented by an education institution as a ``client" and the main asset tested is a Mobile App. Based on that, this section presents a general specification about the performed test, as well as its basic data. In order to organize this set of information, we designed the test process according to the Tramonto framework \cite{DALALANA2017} (see Section \ref{sec:background}), which allows the annotation of all required information for each target scenario.

\subsection{Methodology}

This case study is used as a way to analyze the improvements obtained in a system when using the Tramonto framework. Considering that the tested asset is a mobile application, strategies related to traffic interception, request analysis, investigation of known vulnerabilities and access control attack methods were used. 

\subsection{Fitting Scope}

The following items were set in the Fitting Scope step: Pentest goals, Pentest type, aggressiveness and approach, and other information about dates, estimated time, limitations and constraints. Table~\ref{tab:fittingscope} presents a summary of the Pentest scope.

\begin{table}[h!]
\caption{Pentest summary according to Tramonto}
\begin{tabular}{cc}
\hline
\textbf{Item}         & \textbf{Description}                     \\                           
\hline
\textbf{Client}         & Target University                                                                                                                                                                                                                                  \\ \\
\textbf{Title}          & Target University Mobile App Pentest                                                                                                                                                                                                               \\\\ 
\textbf{Test Description}    & {\begin{tabular}[c]{@{}l@{}}Find vulnerabilities in the mobile app\\\end{tabular}}                                                                                                                                                                                                  \\ \\ 
\textbf{Goals}          & \multicolumn{1}{l}{\begin{tabular}[c]{@{}l@{}}1) Intercept HTTP request and seek sensitive \\ open data;\\\\ 2) Reverse engineer the Android \\ mobile app;\\\\ 3) Find relevant ciphertext or plaintext information\\ in the Android Mobile App;\end{tabular}} \\ \\
\textbf{Date Range}     & 03/30/2019 | 12 hours.                                                                                                                                                                                                                  \\\\ 
\textbf{Type}           & Reversal                                                                                                                                                                                                                                           \\ \\
\textbf{Approach}       & Overt                                                                                                                                                                                                                                              \\ \\
\textbf{Aggressiveness} & High                                                                                                                                                                                                                                               \\ \hline
\end{tabular}
\label{tab:fittingscope}

\end{table}

\subsection{Refinement Tools and Strategies}

By choosing a strategy, according to the Tramonto guidelines, it is possible to give more information about the techniques that should be applied considering the scope definition. From this, the strategy selected can be an alternative to address tasks during the Pentest process - aiming to achieve the goals.

%Apart from the strategy definition, there are the tools.
Usually, several tools are necessary to perform a Pentest. In this case study, the main tools that were used are \texttt{Charles} \cite{charles} and \texttt{Dex2jar} \cite{dex2jar}. \texttt{Charles} is a Web Debugging Proxy tool that enables a developer to view all the HTTP and SSL/HTTPS traffic between a device and the Internet. This includes requests, responses, and HTTP headers (which contain cookies and caching information). \texttt{Charles} was used to find any open request, \textit{i.e}, a request between the app and the server that is not encrypted. This could be the easiest way to discover sensitive data.\texttt{Dex2jar} is a set of tools that handles Android \texttt{.dex} and Java \textit{.class} files. It can be used, \textit{e.g.}, to perform Android APK reverse engineering to retrieve the original source code of \texttt{.apk} files.
%- and also it is the most common way to find.

\section{Case Study - Penetration Execution}
\label{sec:execution}

The Pentest execution, following the Tramonto steps, occurs when the tools and techniques are applied to try the test goals achievement, find vulnerabilities and establish attack vectors. The process started with the requests interception to find open sensitive data or any kind of relevant information. After that, the APK source code was carefully analyzed -  so the decryption/encryption functions were revealed.

The explanation of the activities performed is divided into two steps (detailed in the following subsections): \textbf{Request Intercept} - explaining how the HTTP Requests were intercepted, and \textbf{APK Reverse Engineering} - showing how the reverse engineering process was applied in the APK file to find any sensitive/personal information.

\subsection{Requests Intercept}

The \texttt{Charles} tool was used to intercept the requests and relevant data. Concerning this interception, it was important to set a proper configuration in order to receive and analyze the requests. In this sense, a smartphone was used as a starting point to generate the requests and a proxy was establish from this smartphone. The proxy run on a desktop computer. From that, all HTTP requests were redirect from the Android to the computer. Henceforth, \texttt{Charles} was able to sniff that network traffic.

\begin{figure*}[h]
\centering
\includegraphics[width=0.75\textwidth]{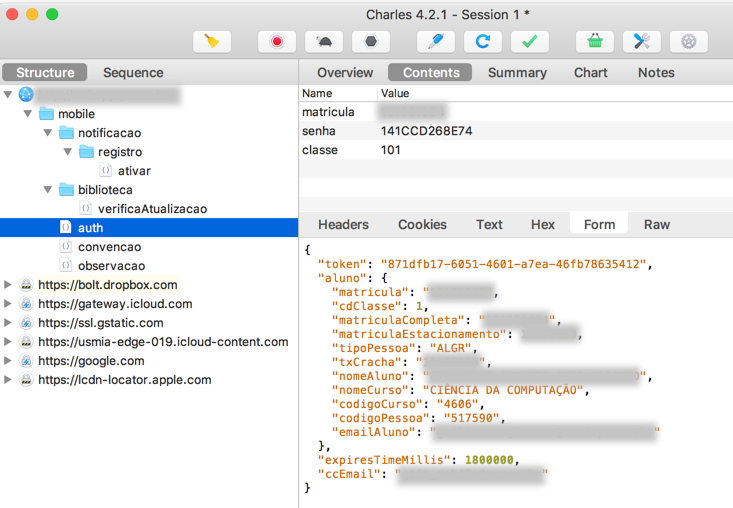}
\caption{User information request intercepted using Charles.}
\label{fig:userinfo}
\end{figure*}

Analyzing one of the open HTTP requests, as presented in Fig.~\ref{fig:userinfo}, the tester is able to retrieve a considerable amount of sensitive data about a specific student - private data was blurred in the figure. This set of information, such as the user \textbf{identification}, \textbf{full name} and \textbf{email} are provided during the requests - that alone is a data breach that should be avoided. 

\begin{table}[htbp]
\caption{Tramonto Criteria}
\begin{center}
\begin{tabular}{cc}
\hline
\textbf{Criteria} & \textbf{Value} \\
\hline
Reproducibility & 8 \\
Impact & 9 \\
Probability & 7 \\
Risk & 6 \\
Priority & 8 \\
\hline

\end{tabular}
\label{tab:tramontoLevels}
\end{center}
\end{table}

The reproducibility level, following Tramonto definitions, can be considered \textbf{high} since this is a simple process to retrieve data. At the same time, the impact level of this attack vector is also high due to the disclosing of sensitive data with almost no effort. By setting the reproducibility and impact levels, Tramonto indicates the Risk, Priority, and Probability values, as shown in Table~\ref{tab:tramontoLevels}.

One simple strategy to mitigate - or even solve - these attack vectors is to use the HTTPS protocol, where the requests are first encrypted to be sent from the app to the server, so that only the destination (that has the decryption key) will be able to read the requests data. Moreover, the adoption of HTTPS could also avoid the sniffing/request interception problem, since the hacker will be able to get the message but not to retrieve its content, as the decryption key is required.

\subsection{APK Reverse Engineering}

As shown in Fig.~\ref{fig:userinfo}, the password field has the value \textbf{141CCD268E74}. Since we knew the university security policy, where the password in the app had exactly four numeric digits, it was possible to conclude that the intercepted password was not in plaintext. The app probably applied a hash function, such as \textit{SHA256}, \textit{WHIRLPOOL} or any other hash function, to the plaintext password before submitting it to the backend application. Thus, the analysis allowed us to know that the password ``encryption'' is being performed through a hash function, which also means that it was performed inside the mobile application. Therefore, analysing the source code of the Mobile App could give us information on which hash function would have been applied to the user password.

APK files are known to be very easy to reverse engineer. \textit{i.e.} given the APK file, it is simple to access the source code from the app through some easy steps - using, for example, the \texttt{dex2jar} tool. This process begins by accessing the app APK file. One way to get the APK file is to download it through an online APK library, such as \texttt{APKPure} \cite{apkpure}.

\begin{figure}
\centering
\includegraphics[width=0.3\textwidth]{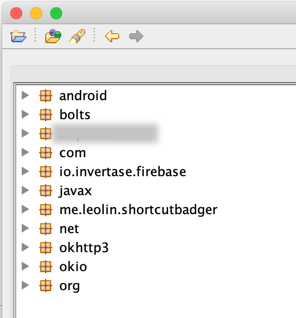}
\caption{Java Decompiler GUI.}
\label{fig:jdgui}
\end{figure}

After that, the \texttt{dex2jar} tool was used to retrieve the source code from the recently downloaded APK - in this example, named \texttt{mobileapp.apk}. The bash code used to decompile the APK was \texttt{\quad d2j-dex2jar.sh -f /mobileapp.apk d2j-de}, which generated a \texttt{.jar} file at the end of the process. Then, a Java decompiler \cite{jdgui} was used to retrieve and export all the source files from the \texttt{.jar} file, as shown in Fig.~\ref{fig:jdgui}. 

\begin{figure*}
\centering
\includegraphics[width=\textwidth]{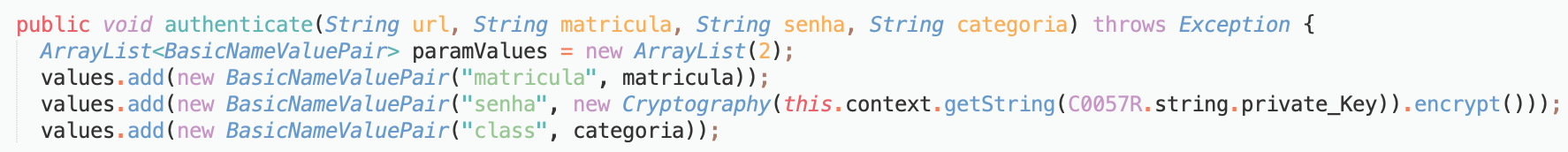}
\caption{Snippet of the authentication function.}
\label{fig:authenticationfunction}
\end{figure*}

By analyzing the code, a function called \texttt{authenticate} inside \texttt{SessionInfo.java} file was found. Hence, this class was explored in order to understand the authentication process. A snippet of the decompiled code is shown in Fig.~\ref{fig:authenticationfunction}.

\begin{figure}
\centering
\includegraphics[width=0.33\textwidth]{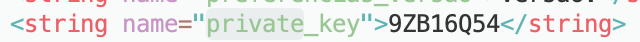}
\caption{Private key.}
\label{fig:privatekey}
\end{figure}

This function shows how the user information is inserted into the HTTP request. Furthermore, it is possible to notice how the password is encrypted — using the \texttt{Cryptography} class. A further analysis showed that the private key was stored inside the \texttt{strings.xml} file. Thus, accessing that file, the private key used to encrypt the user password was found (see Fig.~\ref{fig:privatekey}).

\begin{figure}[h]
\centerline{\includegraphics[width=0.4\textwidth]{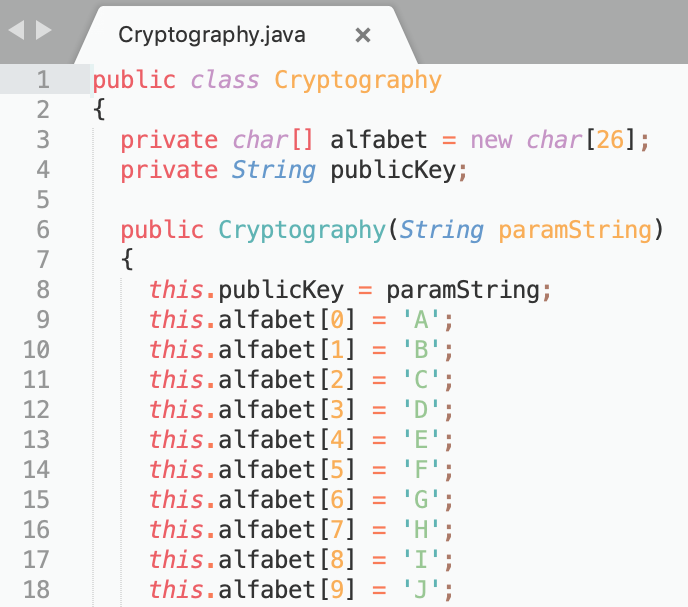}}
\caption{Retrieved Cryptography Java class used in the APK.}
\label{fig:cryptoclass}
\end{figure}

Additionally, by accessing the \texttt{Cryptography} class, we were able to discover that - instead of using known hash functions to encrypt the user password, such as \textit{SHA-2}, \textit{SHA-3} - the app uses a custom implemented hash function, as shown in Fig.~\ref{fig:cryptoclass}. Thus, after a simple analysis of the class function signatures, it is possible to notice that the encryption function is not secure, \textit{i.e.}, the encrypted message can be easily deciphered.

At this point, with the \texttt{Cryptography} class source code and the private key, the test is followed by a task to build a decipher function. Considering the password obtained by the request interception (\textit{141CCD268E74}), it was possible to decipher it easily. The result provided by running the decryption function revealed that the password was \textit{2406}. This password matches exactly the pattern established by the  security policy in the Target University, as mentioned previously.

Taking the discovered student identifier and password to access the portal, it was possible to access the student account. Once authenticated, the hacker has free access to register into classes, and also retrieve financial, academic and other sensitive information that should be protected.

As mentioned before, APK files are easy targets for reverse engineering. In this sense, developers should take care in relation to the data that can be obtained during a request interception or source code analysis. One strategy to improve that could be binary obfuscation \cite{MAVROGIANNOPOULOS2011}.

Finally, the use of a weak custom encryption function should never be used. Nowadays, even secure hash functions (\textit{e.g.}, \textit{SHA-2} and \textit{SHA-3}) may have some weakness when not used properly. For example, there are ``hash crack" bases with most common used passwords. Nonetheless, some best practices can improve security when using hash functions, for example, to use a salt value.

\section{Discussion}
\label{sec:lessons}

Based on the case study presented in this paper, we discuss the usage of Tramonto, Tramonto-app, attack vectors influence on the case study, and some issues regarding data privacy vulnerabilities that must be dealt with by companies:

\begin{itemize}
    \item Tramonto \& Tramonto-App: During the Pentest planning and execution, the use of Tramonto allowed to better conduct the test process. Supported by Tramonto-App, the Tramonto guidelines provide recommendations that helped us to determine what tasks and activities should be performed. At the same time, it was also possible to set vulnerability metrics, to define tools that would be used, to check Pentest goals, to gather the information about the target, and helped to generate a complete report at the end of the Pentest. Besides that, the tasks could be conducted in a more organized way. Furthermore, Tramonto provides ways and alternatives to execute the Pentest, which allowed better management of the Pentest process. Thus, if a tester prefers to use a specific methodology, the Tramonto framework is able to fit the activities according to its structure.
    
    \item Attack Vectors, Techniques and Tools: According to the Tramonto structure, the core of a Pentest is in the Execution step. Hence, this case study showed, in details, the attack vectors used to test the Mobile App. The case study showed also, how a Pentest can be performed in this specific scenario - in a Mobile App assessment. Briefly, the Pentest execution section shows the application of two attack vectors - APK reverse engineering and requests intercept - as well as the usage of popular tools that are adopted when applying these attack vectors. Tha section also introduces the attacker viewpoint when trying to find breaches on a system, and is helpful to understand how to mitigate those breaches and what to do to protect a system against those attack vectors. Hence, we provide, as one of the contributions of this case study, a quick overview of activities and methods commonly used by testers when attacking a mobile app.  
    
    \item Companies vs Vulnerabilities: By describing our case study, we were able to realize that there is no direct relation between the company size and its system vulnerabilities. In our target scenario - an education institution - there is a large set of efforts related to the security concerns and strategies to improve security issues. Considering the business requirements in this specific scenario, we can imagine the amount of personal data handled by the employees and treated by the systems. The privacy aspects, in this sense, require more attention from the managers, considering current issues on privacy regulation, \textit{e.g.} see the European General Data Protection Regulation (GDPR) \cite{gdpr2016}. The emergence of new regulations, laws and compliance criteria are requiring that organizations use structured ways to show that data is protected against breaches. Pentest is a way to provide it and Tramonto may help that.
\end{itemize}

\section{Conclusion and Future Work}
\label{sec:conclusion}

Currently, security issues are producing financial impact in different organisations. Thus, penetration tests are an important step in the life-cycle of modern software development. In order to help the Pentest research field, we presented in this paper the results of a Pentest execution over a Mobile App from a Brazilian university, but this could have been applied to any university that provides similar mobile apps.

The Pentest performed in this paper used the Tramonto framework. Obtained results showed relevant issues in the system and helped software developers to improve the evaluated system. Furthermore, Tramonto provided an organised and guided way to perform Pentests, and helped to identify relevant information from each  executed Pentest and to apply attack vectors.

 Since different companies need some strategy to verify the security of their software, we believe that this paper can contribute in the sense that the case study presented here is a common type of application and its vulnerabilities are also very common. Hence, we believe that the use of frameworks and tools, such as Tramonto, that can assist the tester during the assessments is very important.

As future work, we intend to produce more case studies using different techniques, methods, and tools to help the community with other common mistakes, produced by developers, that can affect the security of an application. Furthermore, we believe that Tramonto is an important way to organize, in a flexible way, Pentests.

% \section*{Acknowledgment}

% This study was financed in part by the Coordenação de Aperfeiçoamento de Pessoal de Nível Superior - Brasil (CAPES) - Finance Code 001 and the Brazilian National Institute of Science and Technology in Forensic Sciences (INCT Forensic Sciences) project (Grant 465450/2014-8). Avelino F. Zorzo is supported by CNPq (grant 315192/2018-6). 

\bibliographystyle{IEEEtran}
\bibliography{pentest} 

\end{document}